\documentclass[%
 reprint,
 amsmath,amssymb,
pra,
floatfix,
]{revtex4-2}

\usepackage{graphicx}
\usepackage{dcolumn}
\usepackage{bm}

\newcommand{\clockT}{$^{1}S_{0}-\,^{3}P_{0}^o$}
\newcommand{\coolingT}{$^{1}S_{0}-\,^{3}P_{1}^o$} 

\begin{document}

\preprint{APS/123-QED}

\title{Isotope shifts for $^1S_0 - {}^3P_{0,1}^o$  Yb lines from multi-configuration Dirac-Hartree-Fock  calculations}

\author{Jesse S. Schelfhout} \email{jesse.schelfhout@uwa.edu.au}
\author{John J. McFerran}%
 \email{john.mcferran@uwa.edu.au}
\affiliation{%
Department of Physics, University of Western Australia, 35 Stirling Highway, 6009 Crawley, Australia
}%




\date{\today}

\begin{abstract}
Relativistic multiconfiguration Dirac-Hartree-Fock (MCDHF) calculations with configuration interaction (CI) are carried out for the $^{1}S_{0}$ and $^{3}P_{0,1}^o$ states in neutral ytterbium by use of the available \textsc{grasp2018} package. From the resultant atomic state functions and the \textsc{ris4} extension, we evaluate the mass and field shift parameters for the $^{1}S_{0}-\,^{3}P_{0}^o$ (clock) and  $^{1}S_{0}-\,^{3}P_{1}^o$ (intercombination) lines. We present improved estimates of the nuclear charge parameters, $\lambda^{A,A'}$, and differences in mean-square charge radii, $\delta\langle r^2\rangle^{A,A'}$, and examine the second-order hyperfine interaction for the $^{3}P_{0,1}^o$ states. Isotope shifts for the clock transition have been estimated by three largely independent means from which we predict the unknown clock line frequencies in bosonic Yb isotopes. Knowledge of these line frequencies has implications for King plot nonlinearity tests and the search for beyond Standard-Model signatures.
\end{abstract}

\maketitle

\section{\label{sec:1}INTRODUCTION}

Atomic systems offer a means to test fundamental physics at a high level of precision in the search for phenomena beyond the Standard Model of elementary particles~\cite{Safronova2018(2),Dzuba2018,Uzan2003}. This may be undertaken by  examining  King plots that are generated  through isotope shift spectroscopy of at least two  transitions in an atomic species~\cite{Counts2020,Berengut2018,Frugiuele2017,Delaunay2017}. Nonlinearities in such  plots may arise due to  higher-order effects within the Standard Model (SM), such as higher-order mass shifts \cite{Yerokhin2020,Shabaev1998}, nuclear deformation~\cite{Allehabi2020,Allehabi2021} or nuclear polarizability~\cite{Flambaum2018}, or due to phenomena beyond the Standard Model \cite{Frugiuele2017,Mikami2017,Delaunay2017,Berengut2018,Flambaum2018,Stadnik2018,Tanaka2020,Debierre2020}. Accurate atomic structure calculations are needed to explore possible causes for such nonlinearities, as is done by investigating additional contributions to isotope shifts beyond the simple mass shift and field shift \cite{Allehabi2021,Berengut2020,Counts2020,Berengut2018,Flambaum2018} --- this can be done by analysing the residuals of a linear fit to a King plot, whereby different nonlinearities are expected to have different signatures in the residuals \cite{Counts2020}. Such calculations can also be used in the extraction of information about the nuclear structure \cite{Reinhard2020,Papoulia2016}.

The recent work of Counts \textit{et al.} \cite{Counts2020}, using narrow optical quadrupole transitions in Yb$^+$, is the only King plot to date which demonstrates nonlinearity beyond the level of experimental uncertainty. This 3$\sigma$ nonlinearity is consistent with interpretations as either higher-order SM effects or physics beyond the SM. Linearity of the Ca$^+$ King plot in Ref. \cite{Solaro2020} suggests that its interpretation as higher-order SM effects should be favoured  \cite{Berengut2020}. Recent work by Allehabi \textit{et al.} \cite{Allehabi2021} suggests that nuclear deformation in Yb nuclei can produce a King plot nonlinearity at a level consistent with that found in \cite{Counts2020}.  A  means of exploring the dominant cause of King plot nonlinearity is by combining prior Yb$^+$ data  with isotope shift measurements of the \clockT\ transition in neutral Yb.  In this work we provide estimates of these \textit{clock} transition frequencies for all the bosonic isotopes of Yb  \textsc{i}, aiding the experimental search for these lines.

Advents in modern computing allow for relativistic atomic structure calculations to be performed with results consistent with experimentally determined values to a few parts in $10^5$ \cite{Rzadkiewicz2018,Koziol2018,Cheng2008,Pasteka2017}. Such computations are also used to determine mass and field shift parameters of isotope shifts for King plot analyses \cite{Kron2020,Counts2020,Solaro2020,Allehabi2020,Kalita2018,Gamrath2018,Berengut2003}. Low-lying energy levels in ytterbium have been explored through computational means \cite{deGroote2020(ArXiv),Dzuba2019,Dzuba2010,Tang2018,Eliav1995,Mani2011,Bostock2011,Porsev1999(2),Kozlov1997,Migdalek1991,Migdalek1986,Liu1998,Kischkel1991,Torbohm1985,Mann1973}, however they usually do not compute isotopes separately and often avoid the ${}^3P_0^o$ state. In this paper, the isotope shifts of the clock and intercombination line (ICL) transitions are computed \textit{ab initio} and the mass and field shift parameters that aid King plot analysis are calculated.

We describe our computational procedure in Sect.~\ref{sec:compmethod}, where the atomic state function is refined through a restricted active set approach using MCDHF-CI computations from a multireference set of configuration state functions. 
Sect.~\ref{sec:IS} summarises the energy level differences and isotope shifts resulting from the MCDHF-CI computations. Sect.~\ref{sec:KF} gives a detailed account of the mass and field shift parameters that are evaluated with \textsc{ris}4 and the calculated atomic state functions. The second order hyperfine interaction is discussed in Sect.~\ref{sec:HFS}, which is necessary to account for the shift in centroid frequencies for the fermionic isotopes.  The nuclear charge parameter is evaluated in Sect.~\ref{sec:nuc} followed by a King plot analysis and estimates of the clock line isotope shifts in Sect.~\ref{sec:King}.  Additional information, including the predictions of the absolute clock line frequencies, is given in the Appendices.

\section{\label{sec:compmethod}COMPUTATIONAL METHOD}

Most \textit{ab initio} isotope shift computations perform computations for a single isotope and then calculate the mass and field shift parameters, using nuclear charge parameter ($\lambda^{A,A'}$) values derived from experiment to arrive at isotope shifts. In contrast, the computations presented here are similar to the ``exact'' method of \cite{Papoulia2016} and to those of \cite{Silwal2020} and \cite{Grant1980}, in that energies and wavefunctions are computed for each isotope of interest, and the isotope shifts are taken as the differences between these energies. It is suggested that this approach can be strongly model-dependent \cite{Ekman2019}, so the more common method of calculating isotope shifts via computed mass and field shift parameters is also pursued in Sections \ref{sec:KF} and \ref{sec:King}.

\begin{figure}
\includegraphics[width=1.05\columnwidth]{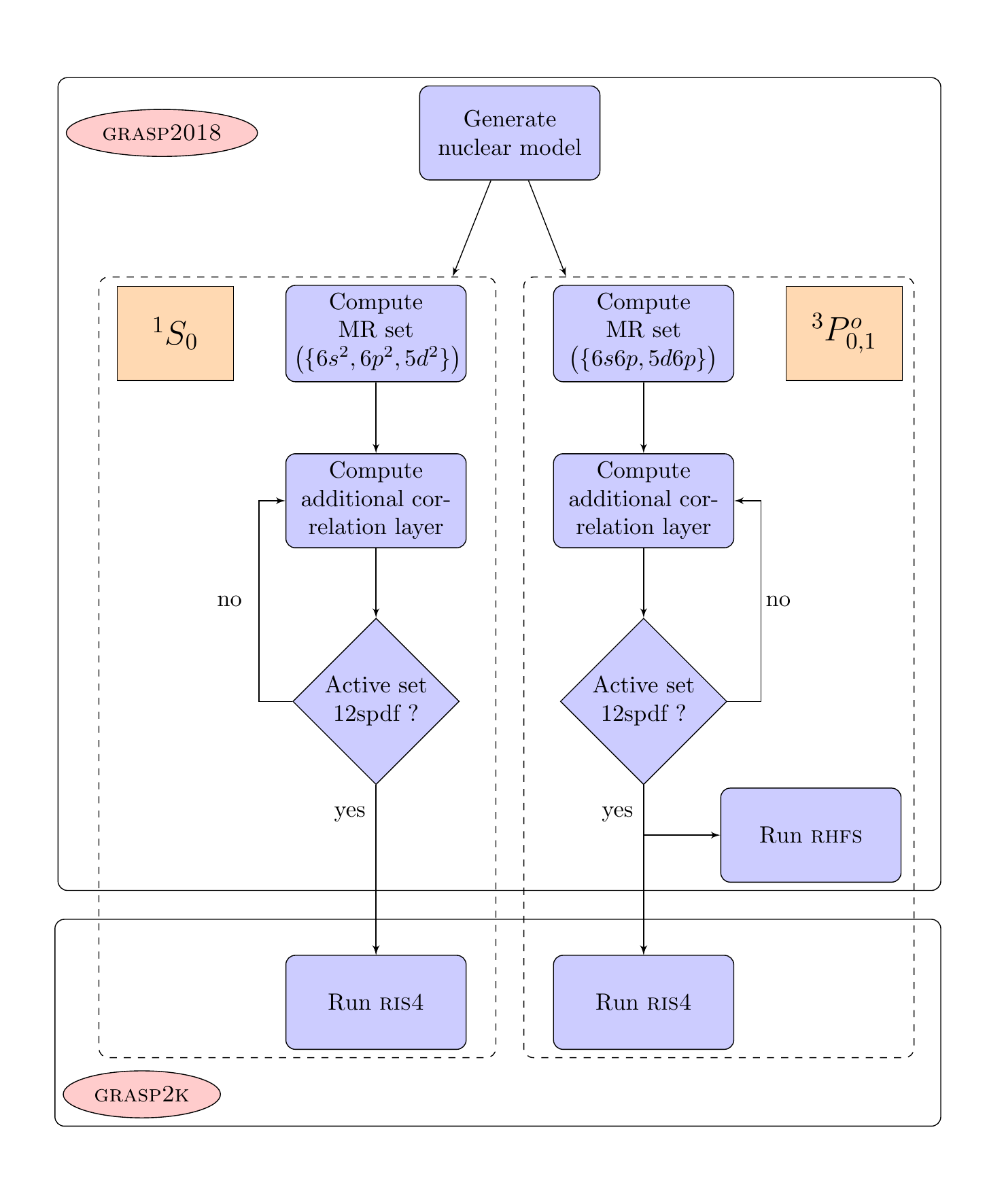}
\caption{\label{fig:method} Procedure for performing MCDHF-CI computations (\textsc{grasp2018}) and extracting isotope shift information (\textsc{grasp2k}). Correlation layers are added until $n=12$ for ground and excited states.  Mass and field shift parameters are evaluated with the \textsc{ris4} package. MR - multireference; \textsc{rhfs} - relativistic hyperfine structure program \cite{Froese_Fischer2018}.  
}
\end{figure}

A two-step approach is used to estimate the isotope shifts, mass shift and field shift parameters for the clock and intercombination transitions using computational methods. First, a MCDHF-CI approach is used to compute the atomic state functions (ASFs) for the ${}^1S_0$ ground state and ${}^3P_{0,1}^o$ excited states using the Fortran 95 package \textsc{grasp2018} (General Relativistic Atomic Structure Package) \cite{Froese_Fischer2018}. Isotope shifts are calculated as the differences in energy between the ground and excited states for different isotopes. Mass and field shift parameters are then extracted using the Fortran 90 program \textsc{ris4} (Relativistic Isotope Shift) \cite{Ekman2019}.  \textsc{ris4} was written as an addition to the \textsc{grasp2k} package \cite{Jonsson2013}; however, we have been able to use it in conjunction with \textsc{grasp2018}\footnote{There is a small incompatibility between \textsc{ris4} and \textsc{grasp2018} in relation to the \texttt{isodata} file, which we overcame by examining the \texttt{isodata} file from \textsc{grasp2k}. \textsc{ris4} is written using the \textsc{grasp2k} architecture and could not be used with \textsc{grasp2018} in isolation (it is currently required to operate inside \textsc{grasp2k}).}.  The computational process is outlined in Figure \ref{fig:method} with further explanation below. A MCDHF-CI approach is used in favour of other approaches, e.g. configuration interaction with many-body perturbation theory (CI+MBPT) \cite{Allehabi2020,Kahl2019}, all-order methods \cite{Safronova2008}, and relativistic coupled cluster (RCC) calculations \cite{Ilyabaev1993}, due to the recent updates of the \textsc{grasp} and \textsc{ris} packages and their cross-compatibility allowing for ease of extraction of isotope shift parameters.

A restricted active space approach is used to construct the atomic state functions, whereby a multi-reference set (MR) is chosen, and additional configuration state functions (CSFs) are systematically included according to allowed substitution rules. The ground state electron configuration for ytterbium is $[\mathrm{Xe}] 4f^{14} 6s^2$. The multi-reference (MR) set for the ${}^1S_0$ ground state is thus taken to be $[\mathrm{Xe}] 4f^{14} \{6s^2,5d^2,6p^2\}$ as these are the configurations with two valence electrons which can form ${}^1S_0$ terms and are near in energy to the $6s^2$ ground state. This is the same MR set as that of the `MCDF IV' approach used in \cite{Migdalek1987}. The excited states ${}^3P_{0,1}^o$ have electron configuration $[\mathrm{Xe}] 4f^{14} 6s 6p$. Conveniently, these can be computed simultaneously using the extended optimal level (EOL) mode of the \textsc{rmcdhf(\_mpi)} program~\cite{Froese_Fischer2018}. Computing the ${}^3P_0^o$ and ${}^3P_1^o$ excited states together with the EOL mode is found to have negligible effect on the clock transition frequency compared with computation of the ${}^3P_0^o$ state on its own ($\sim 0.3\%$ difference). The MR set for  ${}^3P_{0,1}^o$ is taken to be $[\mathrm{Xe}] 4f^{14} \{6s6p,5d6p\}$. The MR sets are summarised in Table \ref{tab:tableMR}, where $N_{CSFs}$ is the number of configuration state functions for the MR set when using relativistic orbital labelling.

\begin{table}
    \caption{\label{tab:tableMR}Multireference configurations for the clock and intercombination transition levels in Yb  \textsc{i}.}
    \begin{ruledtabular}
    \begin{tabular}{cccc}
        Level & $J^{\Pi}$ & MR configurations & $N_\mathrm{CSFs}$\\ \colrule
        $6s^2$ $^1S_0 $ & $0^{+}$ & [Kr]$4d^{10}4f^{14}5s^25p^6+$\{$6s^2,6p^2,5d^2$\} & 5 \\
        $6s6p$ $^3P_0^o$ & $0^{-}$ & [Kr]$4d^{10}4f^{14}5s^25p^6+$\{$6s6p,5d6p$\} & 2 \\
        $6s6p$ $^3P_1^o$ & $1^{-}$ & [Kr]$4d^{10}4f^{14}5s^25p^6+$\{$6s6p,5d6p$\} & 5 \\
\end{tabular}
\end{ruledtabular}
\end{table}

Correlation orbitals are  added layer by layer, where a layer includes 
orbitals for each angular momentum value (e.g. $7s, 7p_-, 7p_+, 6d_-, 6d_+, 5f_-, 5f_+$, with the subscript $\pm$ indicating $j=l\pm 1/2$). Correlation layers are truncated at a principal quantum number of 12~\footnote{The computation falters when trying to compute the radial wavefunction for the $13s$ orbital}.  The new correlation orbitals are optimized using the self consistent field procedure \cite{Froese_Fischer2016} whilst leaving the previously computed orbitals invariant. The correlation layers are built using single and restricted double substitutions from configurations in the MR set. The closed core is taken to be [Kr]. Substitutions from the core are restricted to a single electron from either the $4f_{\pm}$, $5s$ or $5p_{\pm}$ orbitals, together with unrestricted substitutions from the valence orbitals ($6s,6p_{\pm},5d_{\pm}$). This keeps the computations tractable whilst allowing a considerable degree of valence-valence and core-valence correlation. The number of CSFs grows to 30256, 30668 and 84519 for $^1S_0$, $^3P_{0}^o$ and $^3P_{1}^o$, respectively \footnote{This can be compared with the number of CSFs in the MR in Table~\ref{tab:tableMR}}. The dominant CSFs by percentage contribution to the total ASF for each state are tabulated in Appendix~\ref{app:b}. 

The atomic state function computed with all the desired correlation layers is corrected for higher-order QED effects through the \textsc{rci\_mpi} program. The transverse photon interaction is reduced to the Breit interaction by scaling all transverse photon frequencies by a factor of $10^{-6}$~\cite{Froese_Fischer2018}. Vacuum polarisation effects are accounted for, and self-energy is estimated for orbitals up to $n=6$. The normal and specific mass shifts due to the nuclear recoil are also included in the CI computations. Ytterbium nuclei are known to be deformed \cite{Zehnder1975,Clark1979,Allehabi2021}; however, the nuclear model used for these computations --- see Appendix \ref{app:a} --- does not account for nuclear deformation.

\begin{figure}[h]
    \centering
    \includegraphics[width=\columnwidth]{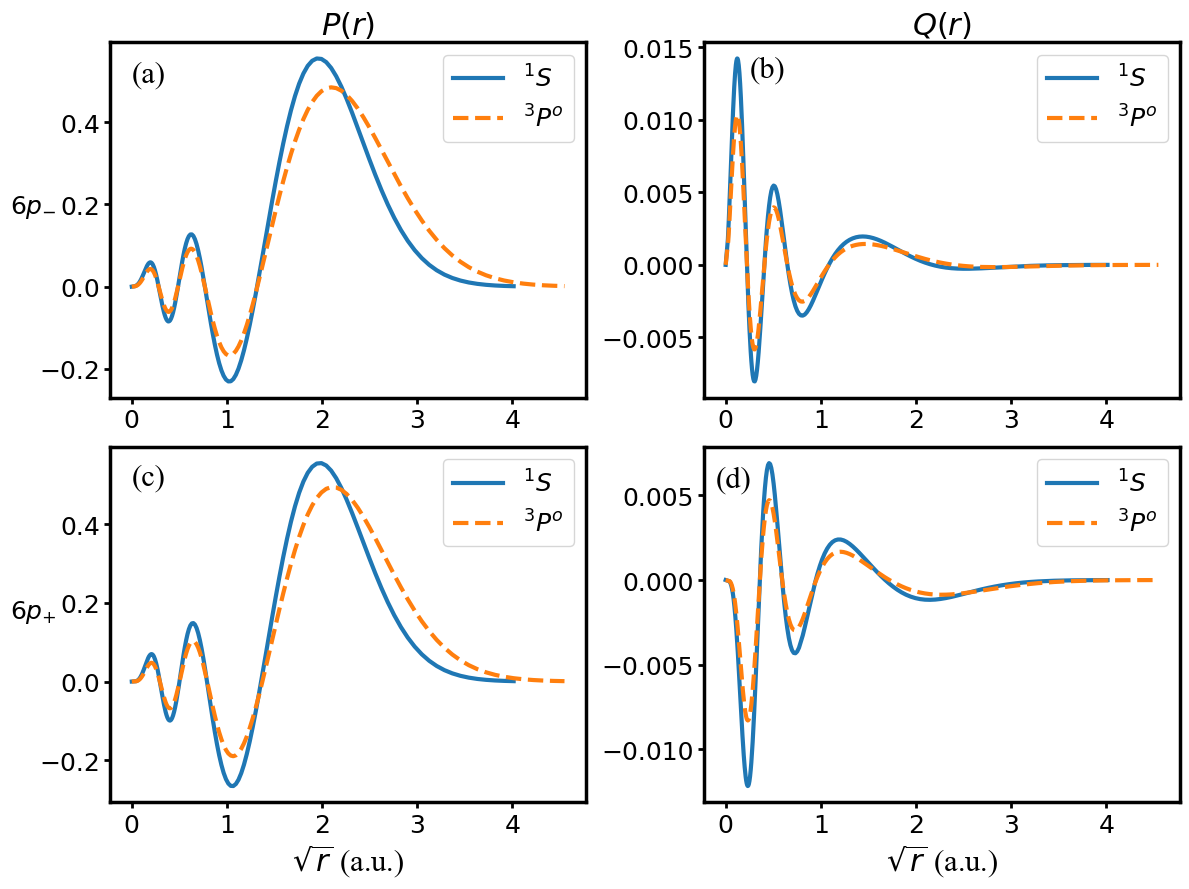}
    \caption{Large, $P(r)$, and small, $Q(r)$, component radial wavefunctions for the ${}^1S_0$ (solid blue) and ${}^3P_{0,1}^o$ (dashed orange) states computed using \textsc{grasp2018}. The abscissa is $\sqrt{r}$ where $r$ is the distance from the centre of the nucleus in atomic units. (a) $P(r)$ for the $6p_-$ orbital. (b) $Q(r)$ for the $6p_-$ orbital. (c) $P(r)$ for the $6p_+$ orbital. (d) $Q(r)$ for the $6p_+$ orbital.}
    \label{fig:rwfnplots}
\end{figure}

The wavefunction arising from a single CSF is an anti-symmetric product of one electron wavefunctions \cite{Grant2007} in the form of a Slater determinant \cite{Slater1929}. The radial functions for the $6p_-$ and $6p_+$ orbitals resulting from our MCDHF-CI computations for the ${}^1S_0$ and ${}^3P_{0,1}^o$ states are represented in Fig.~\ref{fig:rwfnplots}, where the large-component, $P(r)$, and small-component, $Q(r)$, radial functions are presented separately. Less significant deviations between the ground state and excited state radial functions were found for the $6s$ orbital.  

Where possible, uncertainties in the presented computational results are estimated by direct comparison with experiment \cite{Drake2011}. In other cases, uncertainties are estimated by systematically adding correlation layers or increasing the size of the core available for correlation, and analysing the convergence of the desired properties \cite{Froese_Fischer2016}. The latter approach may not include systematic uncertainties arising from the MCDHF-CI method, and so it is desirable to compare against other computational approaches~\cite{Chung2016}. We use a combination of these methods, with quoted uncertainties corresponding to 1$\sigma$ unless otherwise stated.

The efficacy of MCDHF-CI computations has been demonstrated recently; for example, Zhang \textit{et al.} calculate energies for sulfur-like tungsten with near-spectroscopic accuracy \cite{Zhang2020}; Silwal \textit{et al.} compute isotope shifts within the uncertainty bounds of experimental results in Mg-like and Al-like systems~\cite{Silwal2020}; and Palmeri \textit{et al.} produce isotope shifts in reasonable agreement with experimental results for osmium~\cite{Palmeri2016}. Froese Fischer and Senchuk note that good accuracy is generally achieved for light elements or highly-ionised heavy elements, but suggest neutral heavy elements with open core sub-shells can be subject to problems with the accuracy of the calculations or the energies of states not being resolved \cite{Froese_Fischer2020}. These problems are not expected to strongly influence the results of this paper, due to the simple closed-shell electron configuration of neutral ytterbium, in particular the closed $4f^{14}$ sub-shell. Further, neglecting core-core correlations here is justified: The inclusion of core-core correlation in MCDHF computations of neutral lithium and sodium, which have closed core sub-shells, was found to decrease the agreement with experiment compared to computations with only valence and core-valence correlation \cite{Brage1994}. The agreement between computational and experimental values for oscillator strength in singly-ionised thallium, a heavier system than ytterbium with electron configuration $[\mathrm{Xe}]4f^{14}5d^{10}6s^2$, was also found to be better in the absence of core-core correlation \cite{Brage1996}.

The bulk of the computations were performed at the University of Western Australia High Performance Computing Centre on Kaya \footnote{Kaya is a Noongar word meaning `hello'}, one of their high-performance computing machines \footnote{Additional exploratory computations were performed on a laptop with an Intel Core i7-10510U processor with 8 (virtual) cores and 16\,GB of RAM, utilising 6 cores for computations}. Kaya is comprised of two Dell PowerEdge R740 nodes, each with 2 Intel Xeon Gold 6254 processors with 18 cores, 768\,GB of RAM and dual 1.6\,TB NVMe devices. 34 cores were utilised for the computations.

\section{\label{sec:IS}RESULTS: ISOTOPE SHIFTS}  

The computed energy level differences for the clock transition are presented in Table \ref{tab:energy diff clock}, and in Table \ref{tab:energy diff intercombination} for the intercombination line. Energies computed in atomic units ($\mathrm{E_h}$) are converted into frequencies in Hz via multiplication by $2cR_{\infty}=6.579 683 920 502(13) \times 10^{15}\ \mathrm{Hz\ {E_h}^{-1}}$ \cite{Tiesinga2020}. The computed energy level differences are $0.8\%$ larger than the experimental values for the clock transition, and $0.7\%$ larger for the intercombination line.

\begin{table}[h]
    \centering
    \caption{Computed energy level separations and isotope shifts for ${}^1S_0$ ground state and ${}^3P_0^o$ excited state for stable isotopes of Yb. Isotope shifts are relative to ${}^{176}\mathrm{Yb}$.}
    \begin{ruledtabular}
    \begin{tabular}{ccc}
        Isotope & Energy separation (MHz) & Isotope shift (MHz)\\ \colrule
        168 & 522 679 368 & $-5 073$\\
        170 & 522 677 872 & $-3 577$\\
        171 & 522 677 352 & $-3 057$\\
        172 & 522 676 461 & $-2 166$\\
        173 & 522 675 963 & $-1 668$\\
        174 & 522 675 348 & $-1 053$\\
        176 & 522 674 295 & 0
    \end{tabular}
    \end{ruledtabular}
    \label{tab:energy diff clock}
\end{table}

\begin{table}[h]
    \centering
    \caption{Computed energy level separations and isotope shifts for ${}^1S_0$ ground state and ${}^3P_1^o$ excited state for stable isotopes of Yb. Isotope shifts are relative to ${}^{176}\mathrm{Yb}$.}
    \begin{ruledtabular}
    \begin{tabular}{cccc}
        Isotope & Energy separation (MHz) & Isotope shift (MHz)\\ \colrule 
        168 & 543 180 934 & $-5 127$\\ 
        170 & 543 179 422 & $-3 615$\\ 
        171 & 543 178 897 & $-3 090$\\ 
        172 & 543 177 997 & $-2 190$\\ 
        173 & 543 177 493 & $-1 686$\\ 
        174 & 543 176 872 & $-1 065$\\ 
        176 & 543 175 807 & 0 
    \end{tabular}
    \end{ruledtabular}
    \label{tab:energy diff intercombination}
\end{table}

For the clock transition, the isotope shift between ${}^{173}$Yb and ${}^{174}$Yb is calculated as $-615$ MHz and between ${}^{171}$Yb and ${}^{173}$Yb is calculated as $-1389$ MHz. Experimentally these values are found to be $-551.536050(11)$ MHz and $-1259.745595(11)$ MHz, corresponding to percentage differences of $11.5\%$ and $10.3\%$, respectively. Both computed isotope shifts are larger in magnitude than the experimental values. We expect that the isotope shifts presented in Table \ref{tab:energy diff clock} all have an error of approximately $11\%$. The computed isotope shifts for the intercombination line are presented in Table \ref{tab:energy diff intercombination}. These values differ on average by $11.5\%$ when compared with measured values from~\cite{Atkinson2019}. This difference may  reduce with the inclusion of deeper core-valence correlations \cite{Froese_Fischer2018,Gamrath2018,Brage1996}. Variation in nuclear deformation between the isotopes, not accounted for in these computations, may also contribute to the differences between the experimental and computational isotope shifts. While these differences are a concern, they do not prevent us from making viable predictions for clock transition frequencies in the bosonic isotopes (discussed below). The mass and field shift factors determined from these calculations (see Section \ref{sec:KF}) lead to nuclear charge parameters consistent with previous results (see Section \ref{sec:nuc}) and isotope shift estimates consistent with estimates using a method based on experimental results (see Section \ref{sec:King}).

\section{\label{sec:KF}MASS AND FIELD SHIFT PARAMETERS} 
The differences in nuclear mass and charge distributions between isotopes of the same element give rise to small variations in the energy eigenvalues for the atomic system, i.e., isotope shifts. By convention, the isotope shift for a pair of isotopes is calculated by subtracting the energy of the lighter isotope from that of the heavier isotope \cite{King1984}, so for isotopes $A$ and $A'$ with $m_A > m_{A'}$, the isotope shift is given by
\begin{equation}
    \delta\nu^{A,A'} = \nu^A - \nu^{A'}.
\end{equation}

To a very good approximation, an isotope shift may be split into a mass shift and a field shift, arising from differences in the nuclear recoil and nuclear charge distribution, respectively, between the isotopes \cite{King1984}. Under the approximation that the electronic wavefunction for a particular state is invariant between isotopes, the mass and field shifts for an atomic state $i$ factor into electronic and nuclear components
\begin{equation} \label{level isotope shift}
    \delta \nu_i^{A,A'} = K_i \mu^{A,A'} + F_i \lambda^{A,A'},
\end{equation}
where $K_i\ (F_i)$ is the electronic mass (field) shift factor,
\begin{equation}\label{eq:mu}
    \mu^{A,A'} = \frac{1}{m_A} - \frac{1}{m_{A'}} = \frac{m_{A'} - m_A}{m_A m_{A'}}
\end{equation}
is the nuclear mass parameter, and  
\begin{equation} \label{lambda Seltzer}
    \lambda^{A,A'} = \lambda^A - \lambda^{A'} = \sum_{n \geq 1} C_n \delta \langle r^{2n} \rangle^{A,A'}
\end{equation}
is the nuclear charge parameter, where $C_n$ are Seltzer's coefficients \cite{Seltzer1969,Torbohm1985,Blundell1987}. For a transition between an upper state $j$ and a lower state $i$, the isotope shift is given by
\begin{equation} \label{eq:IS1}
    \delta \nu^{A,A'} = \delta \nu_j^{A,A'} - \delta \nu_i^{A,A'} = K \mu^{A,A'} + F \lambda^{A,A'},
\end{equation}
where $K = K_j - K_i$ and $F = F_j - F_i$.

The field shift factor, $F$, has been evaluated for each isotope for both the clock and intercombination transitions with the \textsc{ris4} program following \textsc{grasp2018}. We present the values in Table \ref{tab:IS parameters}, where we see some isotope-dependence. The mean values across all seven stable isotopes are $F_{\mathrm{clock}} = -10.848(21)$ GHz\,fm$^{-2}$ and $F_{\mathrm{ICL}} = -10.951(21)$ GHz\,fm$^{-2}$; we comment on the uncertainties below. For the clock transition, a previously reported value of $F_{\mathrm{clock}}$ was calculated via \textsc{amb}{\scriptsize i}\textsc{t} \cite{Kahl2019}, using configuration interaction only (without MBPT) and a very similar correlation model to this work \cite{Berengut2020}. 
For the intercombination line, the mean value is compared with previous evaluations of $F_{\mathrm{ICL}}$ at the base of the table.  Our value lies approximately central to the range of previous estimations, but with higher precision.

\begin{table}[h]
    \centering
    \caption{Electronic field shift factor ($F$) for the ${}^1S_0 - {}^3P_0^o$ clock transition and the ${}^1S_0 - {}^3P_1^o$ intercombination line (ICL).}
    \begin{ruledtabular}
    \begin{tabular}{ccc}
        Isotope & $F_{\mathrm{clock}}$ (GHz\,fm$^{-2}$) & $F_{\mathrm{ICL}}$ (GHz\,fm$^{-2}$) \\ \colrule
        168  & $-10.865(18)$ & $-10.969(18)$\\
        170  & $-10.855(18)$ & $-10.959(18)$\\
        171  & $-10.852(18)$ & $-10.955(18)$\\
        172  & $-10.846(18)$ & $-10.950(18)$\\
        173  & $-10.843(18)$ & $-10.947(18)$\\
        174  & $-10.839(18)$ & $-10.943(18)$\\
        176  & $-10.833(18)$ & $-10.936(18)$\\ \colrule
        Mean & $-10.848(21)$  & $-10.951(21)$\\
        Ref. \cite{Berengut2020} & $-9.7192$ & ---\\
        Ref. \cite{Fricke2004} & --- & $-9.3(2.1)$\\
        Ref. \cite{Clark1979} & --- & $-10.9$\footnotemark[1]\\
        Ref. \cite{Martensson-Pendrill1994} & --- & $-12.3(0.2)$\\
        Ref. \cite{Jin1991} & --- & $-12.2(0.7)$\footnotemark[1]
    \end{tabular}
    \end{ruledtabular}
    \footnotetext[1]{Value is positive in reference (assumed to be absolute value)}
    \label{tab:IS parameters}
\end{table}

The mass shift factors ($K$) experience little change with isotope; the mean values are $K_{\mathrm{clock}} = -288(75)$\,GHz\,u  and $K_{\mathrm{ICL}} = -280(72)$\,GHz\,u. Note the values are negative. A negative specific mass shift for ${}^3P$ states for two-electron spectra is suggested to arise from angular correlation (private communication in \cite{Migdalek1986}). Whilst these negative mass shifts appear to be at odds with the positive value of $K_{\mathrm{ICL}}=1.5(5)\,\mathrm{THz\,u}$ found by \cite{Martensson-Pendrill1994}, a review of their formulae reveals a difference in sign for the nuclear mass parameter. The same convention ($\mu^{A,A'}>0$ for $m_A>m_{A'}$) is used in \cite{Clark1979,Angeli2013}. The convention used in this work (equation (\ref{eq:mu})) is consistent with that of \cite{Ekman2019,Fricke2004}. Berengut \textit{et al.} \cite{Berengut2020} determine $K_{\mathrm{clock}}=-655$\,GHz\,u using a CI+MBPT method, implemented via \textsc{amb}{\scriptsize i}\textsc{t} \cite{Kahl2019}, demonstrating the dependence of the calculation on the method and supporting its approximate magnitude and sign.

Uncertainties in $K$ and $F$ for each isotope are estimated by systematically increasing the size of the computations. The convergence of the parameters as correlation layers are added for isotope ${}^{174}\mathrm{Yb}$, and as the set of core orbitals available for core-valence correlation is extended for isotope ${}^{176}\mathrm{Yb}$, are presented in Appendix \ref{app:c}. The uncertainties for the mean values (over isotopes) are taken to be the sum by quadrature of (i) the standard deviation of the isotopic data and (ii) the estimated uncertainty for each isotope. The calculated mass shift factor for the intercombination line is consistent with that of \cite{Fricke2004}, and the field shift factor is consistent with \cite{Fricke2004,Clark1979}.

\section{\label{sec:HFS}SECOND-ORDER HYPERFINE STRUCTURE}
The off-diagonal second-order hyperfine interaction for isotopes with nuclear spin results in a shift of the centroid (center of gravity) of the hyperfine manifold relative to that of an isotope with no nuclear spin \cite{Wakasugi1990,Kischkel1991}. Correcting the experimentally-determined centers of gravity for these shifts provides a means of comparison between the bosonic and fermionic isotopes (e.g. for King plot analysis). The shift for a state denoted $|\gamma J I F m_F \rangle$ is given by
\begin{equation} \label{Delta E 2}
    \Delta E_{F}^{(2)} = \sum_{\gamma',J' \neq \gamma,J} \frac{|\langle \gamma J I F m_F | \mathcal{H}_{\mathrm{hfs}} |\gamma' J' I F m_F \rangle|^2}{E_{\gamma,J} - E_{\gamma',J'}}.
\end{equation}
The matrix element in (\ref{Delta E 2}) can be written in terms of the off-diagonal hyperfine structure constants, $A(J,J')$ and $B(J,J')$, as
\begin{widetext}
\begin{align}
    &\langle \gamma' (J-1) I F m_F | \mathcal{H}_{\mathrm{hfs}} | \gamma J I F m_F \rangle = \frac{1}{2} A(J,J-1)\ \sqrt{(K+1)(K-2F)(K-2I)(K-2J+1)} \nonumber\\
    &+ B(J,J-1)\ \frac{[(F+I+1)(F-I)-J^2+1]\sqrt{3(K+1)(K-2F)(K-2I)(K-2J+1)}}{2I(2I-1)J(J-1)},
\end{align}
\end{widetext}
and
\begin{equation}
    \langle \gamma' (J-2) I F m_F | \mathcal{H}_{\mathrm{hfs}} | \gamma J I F m_F \rangle = B(J,J-2),
\end{equation}
where $K=I+J+F$.

Only isotopes ${}^{171}\mathrm{Yb}$ and ${}^{173}\mathrm{Yb}$ have non-zero nuclear spin and thus experience the hyperfine interaction. For $J=1$, the off-diagonal hyperfine constant $B(J,J-1)$ is vanishing. The hyperfine constants calculated using the \textsc{rhfs} program in the \textsc{grasp2018} package \cite{Froese_Fischer2018} and are presented in Table \ref{tab:rhfs}. Uncertainties are taken to be 4\% by comparison of the diagonal hyperfine constants with the experimental values from Atkinson \textit{et al.} \cite{Atkinson2019}.
\begin{table}[h]
    \centering
    \caption{Hyperfine interaction constants calculated using \textsc{rhfs}.}
    \begin{ruledtabular}
    \begin{tabular}{cccc}
        Isotope & $A({}^3P_1)$ (GHz) & $A({}^3P_1,{}^3P_0)$ (GHz) & $B({}^3P_1)$ (GHz) \\ \colrule
        171 & 4.07(17) & 3.89(16) & 0 \\
        173 & -1.12(5) & -1.07(5) & -0.794(32)
    \end{tabular}
    \end{ruledtabular}
    \label{tab:rhfs}
\end{table}

Calculation of the centroid shift using (\ref{Delta E 2}) makes use of the energy difference between the fine-structure levels, ${}^3P_0^o$  and ${}^3P_1^o$; i.e., the value of 21 092 574.882(93) MHz for  ${}^{174}\mathrm{Yb}$, based on measurements presented in \cite{Atkinson2019} and \cite{Poli2008}. The centroid shifts for the clock transition, to second-order in perturbation theory, for the mixing of the ${}^3P_0$ and ${}^3P_1$ states we calculate to be $-0.537(44)$\,MHz and $-0.476(39)$\,MHz for ${}^{171}\mathrm{Yb}$ and ${}^{173}\mathrm{Yb}$, respectively. For the ICL, the $F=I$ hyperfine level is the only one influenced by mixing with the ${}^3P_0^o$ state, and so the shift to its centroid is smaller. The new centroids for the ICL isotope shifts relative to ${}^{176}\mathrm{Yb}$ are $-1510.948(42)$\,MHz for ${}^{173}\mathrm{Yb}$, and $-2781.182(54)$\,MHz for ${}^{171}\mathrm{Yb}$ (\textit{c.f.} Ref. \cite{Atkinson2019}).

The centers of gravity for the intercombination line isotope shifts presented in \cite{Atkinson2019} are correct to first-order in perturbation theory; however, the second-order corrections due to mixing with the ${}^3P_0^o$ state are greater than the experimental uncertainty and so are accounted for here. The effects of mixing with other nearby states (${}^1P_1^o,{}^3P_2^o$) are estimated to be less than experimental uncertainty. The centers of gravity determined from the measured clock transition frequencies for ${}^{171}\mathrm{Yb}$ \cite{McGrew2019,Pizzocaro2020} and ${}^{173}\mathrm{Yb}$ \cite{Clivati2016} must also take into account the higher-order perturbations in order to make comparison with that of ${}^{174}\mathrm{Yb}$ \cite{Poli2008} in a King plot analysis. The resultant isotope shifts (between centers of gravity for the fermions) are presented in Table \ref{tab:corrected isotope shifts}.  The values are used later in Sect.~\ref{sec:King}.

\begin{table}[h]
    \centering
    \caption{Isotope shifts for the ${}^1S_0 - {}^3P_1^o$ intercombination line and ${}^1S_0 - {}^3P_0^o$ clock line in Yb \textsc{i}. $\delta\nu^{A,A'} = \nu^A - \nu^{A'}$. The centroid for the hyperfine manifold is used for fermionic isotopes, where the corrections to second order are taken into account.}
    \begin{ruledtabular}
    \begin{tabular}{cccc}
        $A$ & $A'$ & $\delta\nu_{\mathrm{ICL}}^{A,A'}$ (MHz) & $\delta\nu_{\mathrm{clock}}^{A,A'}$ (MHz)\\ \colrule
        176 & 174 & $-954.734(31)$ & ---\\
        174 & 172 & $-1000.792(48)$ & ---\\
        172 & 170 & $-1285.816(81)$ & ---\\
        170 & 168 & $-1369.602(93)$ & ---\\
        173 & 172 & $-444.578(56)$ & ---\\  
        172 & 171 & $-825.656(65)$ & ---\\  
        171 & 170 & $-460.160(91)$ & ---\\  
        174 & 173 & $-556.214(53)$ & $-552.012(39)$\\  
        173 & 171 & $-1270.234(69)$ & $-1259.807(58)$  
    \end{tabular}
    \end{ruledtabular}
    \label{tab:corrected isotope shifts}
\end{table}

\section{\label{sec:nuc}NUCLEAR CHARGE PARAMETER} 

The nuclear charge parameter can be calculated by rearranging equation (\ref{eq:IS1}) to find,
\begin{equation} \label{nuclear parameter}
    \lambda^{A,A'} = \frac{1}{F} \left( \delta \nu^{A,A'} - K \mu^{A,A'} \right).
\end{equation}
By use of the isotope shifts presented in Table \ref{tab:corrected isotope shifts}, the mass shift and field shift parameters calculated in Table \ref{tab:IS parameters}, and the isotope masses presented in Appendix \ref{app:a}, the nuclear charge parameter {$\lambda^{A,A'}$} can be determined from Eq.~\ref{nuclear parameter}, as presented in Table \ref{tab:nuclear parameters}. The uncertainties are dominated by the uncertainty in $K$, but they are lower than those of previous estimates by at least a factor of four. King \cite{King1984} notes that the values from Clark \textit{et al.} \cite{Clark1979} give excessive weight to the muonic and x-ray data in their combined analysis, which leads to  larger values than our own. Column 5 shows $\lambda^{A,A'}$ values from Clark \textit{et al.} based on optical data alone, showing better agreement with our values. Jin \textit{et al.} \cite{Jin1991} assume a specific mass shift of zero and use a larger value for the field shift parameter (12.2\,GHz\,fm$^{-2}$), leading to their lower values for $\lambda^{A,A'}$.

\begin{table*}[h]
    \centering
    \caption{Nuclear charge parameters {$\lambda^{A,A'}$}  determined from  the Yb \textsc{i} intercombination line measurements and calculated $F$ parameters $-$  in units of $10^{-3}\,\mathrm{fm}^2$ (column 3). Data from prior works are presented for comparison.}
    \begin{ruledtabular}
    \begin{tabular}{ccccccc}
        $A$ & $A'$ & This work & Ref. \cite{Clark1979}\footnote{Combined analysis of optical, x-ray \& muonic isotope shifts} & Ref. \cite{Clark1979}\footnote{Optical isotope shifts only} & Ref. \cite{Jin1991} & Ref. \cite{Martensson-Pendrill1994} \\ \colrule
        176 & 174 & 88.86(47) & 109(8) & 87(13) & 79.4(4.0) & 86(2)\\
        174 & 172 & 93.10(48) & 114(8) & 92(15) & 83.3(4.2) & 90(2)\\
        172 & 170 & 119.17(51) & 139(8) & 116(16) & 106.6(5.3) & 113(3)\\
        170 & 168 & 126.86(53) & 147(8) & 128(19) & 113.6(5.7) & 120(14)\footnotemark[3] \\
        173 & 172 & 41.46(24) & 53(4) & 41(10) & 37.1(1.9) & 40(1)\\
        172 & 171 & 76.27(27) & 85(4) & --- & 68.3(3.4) & 71(1)\\
        171 & 170 & 42.90(25) & 54(4) & 41(10) & 38.3(1.9) & 42(1)\\
        174 & 173 & 51.64(25) & 61(4) & --- & 46.2(2.3) & 49(1)\\
        173 & 171 & 117.72(50) & --- & --- & --- & 110(2)\footnotemark[3]
    \end{tabular}
    \end{ruledtabular}
    \footnotetext[3]{Value calculated using results from Ref. \cite{Martensson-Pendrill1994}}
    \label{tab:nuclear parameters}
\end{table*}

The nuclear charge parameter, $\lambda^{A,A'}$ can be converted into the difference in mean-square nuclear charge radii, $\delta\langle r^2\rangle^{A,A'}$, through rescaling \cite{Martensson-Pendrill1994,Fricke2004} or using an iterative procedure \cite{Angeli2004,Angeli2013}. Fricke and Heilig \cite{Fricke2004} determine the higher-order moments to contribute $-5.9\,\%$ to $\lambda^{A,A'}$ based on experimental data from muonic atoms, so the differences in mean-square charge radii are recovered in this work by rescaling via $\delta\langle r^2\rangle^{A,A'} = \lambda^{A,A'}/0.941$. Table \ref{tab:delta r squared} presents the differences in mean-square charge radii arising from this work and previous works. The tabulated $\delta\langle r^2\rangle^{A,A'}$ values for Yb in Angeli and Marinova \cite{Angeli2013} are calculated using semi-empirical mass shift and field shift parameters of $F_{\mathrm{ICL}}=-11.5\,\mathrm{GHz\,fm}^{-2}$ and $K_{\mathrm{ICL}}=-4.6(1.6)\,\mathrm{THz\,u}$ \footnote{The sign of this value has been corrected for the difference in sign convention between Ref. \cite{Angeli2013} and this work}. This mass shift parameter is much larger in magnitude than that calculated in this work and by \cite{Berengut2020}, leading to the tabulated values being larger than those determined in this work. Allehabi \textit{et al.} \cite{Allehabi2021} also suggest that the tabulated $\delta\langle r^2\rangle^{A,A'}$ values are too large based on their own nuclear and electronic structure calculations.

\begin{table*}[h]
    \centering
    \caption{Differences in mean-square nuclear charge radii {$\delta\langle r^2\rangle^{A,A'}$}, determined from the $\lambda^{A,A'}$ values in Table \ref{tab:nuclear parameters} via $\delta\langle r^2\rangle^{A,A'} = \lambda^{A,A'}/0.941$ \cite{Fricke2004}, in units of $10^{-3}\,\mathrm{fm}^2$ (column 3). Data from other works are presented for comparison.}
    \begin{ruledtabular}
    \begin{tabular}{cccccccc}
        $A$ & $A'$ & This work & Ref. \cite{Jin1991} & Ref. \cite{Martensson-Pendrill1994} & Ref. \cite{Angeli2013}\footnote{Ref. \cite{Angeli2013} presents only statistical errors in the uncertainty --- the large uncertainty in the mass shift parameter used in calculation is not propagated through. Propagating the uncertainty from the mass shift parameter leads to an uncertainty of $\sim9\times10^{-3}\,\mathrm{fm}^2$ for the first row.} & Ref. \cite{Fricke2004} & Ref. \cite{Allehabi2021}\footnote{Purely computational values (presented without uncertainty)}\\ \colrule
        176 & 174 & 94.4(0.5) & 84.8(4.6) & 90(2) & 115.9(0.1) & 114(30) & 97\\
        174 & 172 & 98.9(0.6) & 88.8(4.6) & 95(3) & 120.7(0.2) & 118(28) & 102\\
        172 & 170 & 126.6(0.6) & 113.5(6.6) & 119(4) & 147.9(0.2) & 151(36) & 130\\
        170 & 168 & 134.8(0.6) & 121.0(7.2) & 125(15) & 156.1(0.4) & 160(126) & 138\\
        173 & 172 & 44.1(0.3) & --- & 42(2) & 55.6(0.2) & 52(19) & ---\\
        172 & 171 & 81.1(0.3) & --- & 75(2) & 90.7(0.2) & --- & ---\\
        171 & 170 & 45.6(0.3) & --- & 44(2) & 57.2(0.2) & 55(80) & ---\\
        174 & 173 & 54.9(0.3) & --- & 51(2) & 65.1(0.2) & --- & ---\\
        173 & 171 & 125.1(0.6) & --- & 116(3) & 146.3(0.2) & --- & ---
    \end{tabular}
    \end{ruledtabular}
    \label{tab:delta r squared}
\end{table*}

\section{\label{sec:King}KING PLOT AND CLOCK TRANSITION ISOTOPE SHIFTS}

\begin{figure}[h]
    \includegraphics[width=\columnwidth]{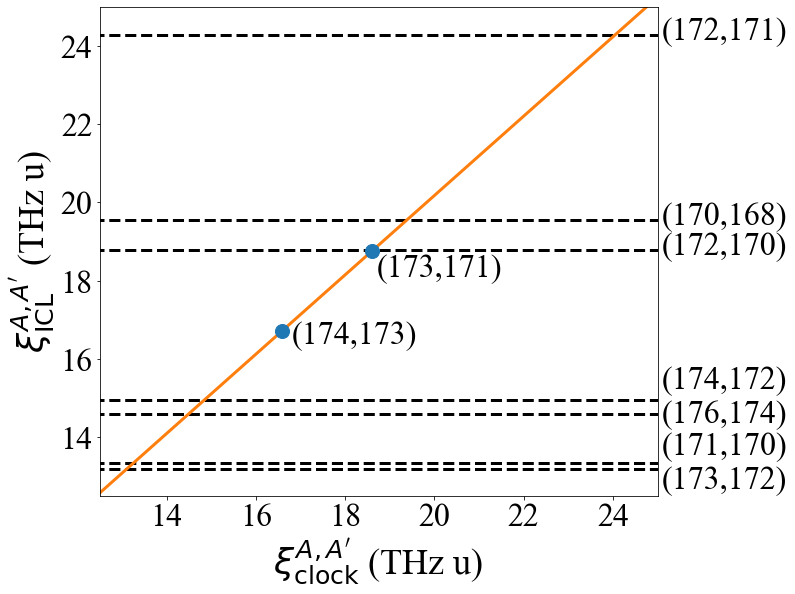}
    \caption{  King plot for clock and intercombination lines of Yb \textsc{i}. Blue circles represent isotope pairs $(A,A')$ with clock transition measurements. Dashed black lines represent isotope pairs $(A,A')$ without clock transition measurements. The solid orange line is the King linearity relationship. 
    Error bars are smaller than the marker size.} 
    \label{fig:King plot}
\end{figure}

A King plot compares the isotopic shifts of one transition, $i$, against that of another, $j$. By scaling the isotope shift with the reciprocal of the nuclear mass parameter, one defines the \textit{modified} isotope shift, 
\begin{equation}
   \xi_i^{A,A'} =\delta\nu_i^{A,A'}/\mu^{A,A'}.
\end{equation}
From Eq.~\ref{eq:IS1} and assuming the nuclear parameters $\lambda^{A,A'}$ and $\mu^{A,A'}$ are the same for both (all) transitions, one finds,
\begin{equation}
  \xi_i^{A,A'} = (F_i/F_j)\xi_j^{A,A'} +(K_i-K_jF_i/F_j).
\end{equation}
A plot of $\xi_i^{AA'}$ versus $\xi_j^{AA'}$  should thus, to first order, form a straight line with slope $F_i/F_j$ and intercept $K_i-K_jF_i/F_j$, known as a King plot. A King plot constructed from the measured isotope shifts for the clock and intercombination lines in Yb \textsc{i} is presented in Figure \ref{fig:King plot}. With only three isotopes having frequency measurements for the clock transition, only two independent data points can be used to create the King plot (the 171-174 pairing makes it overdetermined). The gradient and intercept for the linear `fit' are 1.0138(12) and $-0.102(22)$\,THz\,u, respectively, where the uncertainties are derived from the square roots of the diagonal entries to the covariance matrix calculated using an orthogonal distance regression \cite{Boggs1990}.

The gradient is given as $F_{\mathrm{ICL}} / F_{\mathrm{clock}}$ and the intercept as $K_{\mathrm{ICL}} - K_{\mathrm{clock}} F_{\mathrm{ICL}} / F_{\mathrm{clock}}$. The calculated values in Table \ref{tab:IS parameters} produce a gradient of 1.0095(28), and an intercept of 0.01(11)\,THz\,u. This gradient and intercept values are not inconsistent with those obtained from the King plot in Fig.~\ref{fig:King plot} (experimental).

The unknown isotope shifts for the ${}^1S_0 - {}^3P_0^o$ clock transition  can be estimated  in three different ways (further information follows), 

\begin{enumerate}
    \item Energy level differences for each isotope are found through MCDHF-CI computations.  From these, the isotope shifts are evaluated, and because there is a consistent offset from measured values in \coolingT, can be scaled to match the three known experimental isotope shifts.  
    \item The mass shift and field shift parameters calculated using \textsc{ris4} are used with the ICL isotope shifts presented in Table \ref{tab:corrected isotope shifts} to estimate the clock transition isotope shifts. This estimate is based predominantly on theoretical calculation.
    \item The modified frequency shifts are extrapolated from a King plot constructed using the clock and intercombination lines, and converted back into isotope shifts. This estimate is based predominantly on experimental measurement.
\end{enumerate}
The estimated isotope shifts for the clock transition for each method are presented in Table \ref{tab:clock isotope shift estimates}.

\begin{table}[h]
    \centering
    \caption{Clock transition isotope shifts ($\delta\nu^{A,A'}_{\mathrm{clock}}$) in MHz determined using three different methods, as outlined in the text. 
    }
    \begin{ruledtabular}
    \begin{tabular}{ccccc}
       $A$ & $A'$ & Method-1 & Method-2 & Method-3 \\ \colrule
        176 & 174 & $-949(10)$ & $-945.1(7.3)$ & $-949.5(2.8)$\\
        174 & 172 & $-1002(11)$ & $-990.7(7.5)$ & $-995.0(2.9)$\\
        172 & 170 & $-1272(13)$ & $-1273.0(8.0)$ & $-1275.3(3.3)$\\
        170 & 168 & $-1347(14)$ & $-1356.0(8.2)$ & $-1357.9(3.4)$\\
        173 & 172 & $-448(5)$ & $-440.0(3.7)$ & $-443.0(1.4)$\\
        172 & 171 & $-803(9)$ & $-817.5(4.2)$ & $-816.7(2.0)$\\
        171 & 170 & $-469(5)$ & $-455.5(3.8)$ & $-458.5(1.5)$\\
        174 & 173 & $-554(6)$ & $-550.6(3.8)$ & $-552.0(1.5)$\\
        173 & 171 & $-1251(13)$ & $-1257.6(7.9)$ & $-1259.8(3.3)$
    \end{tabular}
    \end{ruledtabular}
    \label{tab:clock isotope shift estimates}
\end{table}

(Method-1): The \textit{ab initio} isotope shifts calculated for the clock transition using MCDHF-CI computations, presented in Table \ref{tab:energy diff clock}, are larger than experimental values  by  $\sim11$\% (for all the isotopes). This difference we attribute to a systematic effect in the calculations, which we can account for by a  scaling factor. Accounting for the difference leads to the estimates given in the `Method-1' column of Table \ref{tab:clock isotope shift estimates}. The adjusted isotope shift between ${}^{173}$Yb and ${}^{174}$Yb is $-554$ MHz and between ${}^{171}$Yb and ${}^{173}$Yb is $-1251$ MHz, at differences from experiment of $0.5\%$ and $-0.7\%$, respectively. In line with these values we place uncertainties of 1\% on the remaining shifts in Table~\ref{tab:clock isotope shift estimates} (Method-1).  We regard this as the least reliable estimate of the unmeasured clock line isotope shifts.

(Method-2): Equation \ref{eq:IS1} applies for both the clock and ICL transitions, with the nuclear parameters taken to be independent of the electronic states. Substituting for $\lambda^{A,A'}$ between these two equations leads to
\begin{equation}
    \delta\nu_{\mathrm{clock}}^{A,A'} = \left( K_{\mathrm{clock}} - \frac{F_{\mathrm{clock}}}{F_{\mathrm{ICL}}} K_{\mathrm{ICL}} \right) \mu^{A,A'} + \frac{F_{\mathrm{clock}}}{F_{\mathrm{ICL}}} \delta\nu_{\mathrm{ICL}}^{A,A'}.
\end{equation}
The ICL isotope shifts presented in Table \ref{tab:corrected isotope shifts} can be used with the calculated mass shift and field shift parameters to arrive at the clock transition isotope shifts. This is equivalent to constructing a King plot using the theoretical mass and field shifts computed using \textsc{ris4} and nuclear charge parameters presented in Table \ref{tab:nuclear parameters}, and leads to the isotope shifts presented in the `Method-2' column of Table \ref{tab:clock isotope shift estimates}. The uncertainties are again dominated by the uncertainties in the $K$ parameters for each transition, similarly to those for Table \ref{tab:nuclear parameters}.

(Method-3):  Assuming King linearity holds, the King plot in Figure \ref{fig:King plot} can be extrapolated to arrive at the clock transition isotope shifts for other pairings. These estimates are presented in the final column of Table \ref{tab:clock isotope shift estimates}. We emphasize that the King plot is based on experimental values and not MCDHF-CI calculations. The only computational component is that of the higher order hyperfine shifts affecting the centers of gravity. Consistent with this, the uncertainties for `Method-3' are less than those of `Method-1'. The values in the final two rows of this column provide a consistency check, since these are the isotopes used to construct the King plot --- they agree within the uncertainties. For comparison, the experimental values appear in Table \ref{tab:corrected isotope shifts}. The presented uncertainties for `Method-3' are calculated using propagation of errors with the uncertainties from the ICL isotope shifts, nuclear masses, and fit parameters.

The isotope shifts for Methods 2 and 3 presented in Table~\ref{tab:clock isotope shift estimates} provide a region in which experimental searches can be made for the bosonic clock transitions. A weighted mean of the shifts has been used to estimate the absolute clock transition frequencies, as tabulated in Appendix~\ref{app:e}.

\section{\label{sec:conclusion}CONCLUSIONS}
\textit{Ab initio} computations of the isotope shifts for the clock transition and its partnering intercombination line (${}^1S_0 - {}^3P_1^o$) have been performed separately for each stable isotope using a MCDHF-CI method implemented by the \textsc{grasp2018} \cite{Froese_Fischer2018} package. Absolute transition frequency measurements agree with experimental results to less than 1\% error, with isotope shifts differing from experimental values by 11\%. Using these same computations, the hyperfine interaction constants for the ${}^3P_1^o$ state have been calculated to within 4\% of corresponding experimental values. Corrections of the centroids of the hyperfine manifolds for the second-order hyperfine interaction in the fermionic isotopes have also been made.

The electronic mass shift and field shift parameters are computed with the program \textsc{ris4} \cite{Ekman2019} using the results of the MCDHF-CI computations. The corrected isotope shifts for the intercombination line together with these electronic mass shift and field shift parameters enable computation of the nuclear charge parameters, $\lambda^{A,A'}$, consistent with previous results, but with an estimated order of magnitude reduction in uncertainties. The differences in mean-square charge radii, $\delta\langle r^2\rangle^{A,A'}$, are calculated and found to be significantly smaller than tabulated values in Angeli and Marinova \cite{Angeli2013}.

Experimental isotope shifts for the clock and intercombination lines, corrected for the second-order hyperfine interaction, have been used to construct a King plot with two data points. This King plot is used to estimate the isotope shifts for the clock transition for the undiscovered bosonic isotopes. These estimates are found to be reasonably consistent with estimates based on the calculated mass shift and field shift parameters.

The computations may be increased in size by including deeper core-valence correlation, and by extending the active set of orbitals beyond a principal quantum number of 12, given sufficient computational resources. The inclusion of deeper core-valence correlation is expected to reduce the 11\% discrepancy between the computed and experimental isotope shifts \cite{Froese_Fischer2018,Gamrath2018}. Different nuclear models, including models accounting for the known deformation of Yb nuclei, may also be explored to investigate their potential systematic effects on the computed results.

With suggestions to combine the results of Counts \textit{et al.} \cite{Counts2020} with isotope shift measurements of a clock transition in neutral ytterbium \cite{Counts2020,Berengut2020}, the undiscovered bosonic-isotope clock transitions should be sought using the isotope shift estimates presented in this work (e.g. with cold Yb atoms in an optical lattice trap and a DC magnetic field applied~\cite{Barber2006}). Once the clock isotope shifts are identified, King plots can be constructed with other high-precision isotope shift measurements in neutral and ionised ytterbium in order to investigate King nonlinearity and identify or constrain physics beyond the Standard Model.

\begin{acknowledgments}
We are grateful for the assistance provided by Christopher Bording and Hayden Walker from the UWA High Performance Computing Team. J.~S. acknowledges support from the University of Western Australia's Winthrop Scholarship, and St Catherine's College. This research was undertaken with the assistance of resources from the University of Western Australia High Performance Computing Team.
\end{acknowledgments}

\appendix

\section{\label{app:a} NUCLEAR MODEL}
The nuclear charge distribution is modelled as a two-component Fermi distribution \cite{Martensson-Pendrill2003,Hahn1956}
\begin{equation}
    \rho(r) = \frac{\rho_0}{1+e^{(r-c)/a}},
\end{equation}
where $c$ is the half-density radius, $a$ is related to the nuclear skin thickness $t$ by $t=(4 \ln{3}) a$, and $\rho_0$ is a normalisation factor such that
\begin{equation}
    \int_{0}^{\infty} 4\pi r^2 \rho(r) dr = Z.
\end{equation}
For all isotopes the atomic number is $Z=70$ and the nuclear skin thickness is taken to be $t=2.18(2)$ fm \cite{Zehnder1975}. This value for the nuclear skin thickness is less than the typical value of $t=2.3$ fm assumed for most nuclei \cite{Martensson-Pendrill2003}; however, it is the only value found for Yb which includes an explicit uncertainty. Other authors have used $t=2.3$ fm \cite{Martensson-Pendrill1994} or $t=2.4$ fm \cite{Shorifuddoza2019}. The dependence of the results upon the skin thickness was investigated and found to be insignificant. The nuclear parameters used in the MCDHF-CI computations are presented in Table \ref{tab:nuclear model}. In addition to these, the only isotopes with non-zero nuclear spin and magnetic dipole moment are ${}^{171,173}\mathrm{Yb}$. ${}^{171}\mathrm{Yb}$ has a nuclear spin of $I=1/2\,\hbar$ and a magnetic dipole moment of $\mu = 0.49367(1)\,\mu_N$ \cite{Stone2005}. ${}^{173}\mathrm{Yb}$ has a nuclear spin of $I=5/2\,\hbar$, a magnetic dipole moment of $\mu = -0.67989(3)\,\mu_N$ \cite{Stone2005}, and nuclear electric quadrupole moment of $Q=2.80(4)\,\mathrm{b}$ \cite{Zehnder1975}.

\begin{table}[h]
    \centering
    \caption{Isotope-dependent parameters for the Yb nuclear model. $A$, mass number; $R$, rms nuclear charge radius; $m$, atomic mass. $R$ values obtained from \cite{Angeli2013}, and mass values are obtained from \cite{Wang2017} except for ${}^{168}\mathrm{Yb}$ \cite{Nesterenko2020}.}
    \begin{ruledtabular}
    \begin{tabular}{ccc}
        $A$ & $R$ (fm) & $m$ (u)\\
        \colrule
        168 & 5.2702(56) \cite{Angeli2013} & 167.93389132(10) \cite{Nesterenko2020}\\
        170 & 5.2853(56) \cite{Angeli2013} & 169.934767246(11) \cite{Wang2017}\\
        171 & 5.2906(57) \cite{Angeli2013} & 170.936331517(14) \cite{Wang2017}\\
        172 & 5.2995(58) \cite{Angeli2013} & 171.936386659(15) \cite{Wang2017}\\
        173 & 5.3046(59) \cite{Angeli2013} & 172.938216215(12) \cite{Wang2017}\\
        174 & 5.3108(60) \cite{Angeli2013} & 173.938867548(12) \cite{Wang2017}\\
        176 & 5.3215(62) \cite{Angeli2013} & 175.942574709(16) \cite{Wang2017}
    \end{tabular}
    \end{ruledtabular}
    \label{tab:nuclear model}
\end{table}

\section{\label{app:b} STATE COMPOSITIONS}
The atomic state functions determined using the MCDHF-CI method consist of weighted combinations of many configuration state functions (CSFs). The percentage contributions of the most significant CSFs are listed for the ${}^1S_0$ ground state and the ${}^3P_{0,1}^o$ excited states in Table \ref{tab:CSFs jj}. Our values are consistent with those reported by Migdalek and Baylis~\cite{Migdalek1991}, where their calculation extended only to our MR set.

\begin{table}[h]
    \centering  
    \caption{The highest contributing CSFs in the compositions of three Yb \textsc{i} atomic states.} 
    \begin{ruledtabular}
    \begin{tabular}{lr}
        CSF & Percentage\\ \colrule
        \multicolumn{2}{c}{${}^1S_0$}\\ \colrule
        $6s^2$ & 91.56\%\\
        $6p_+{\hspace{-0.5ex}}^2$ & 1.87\%\\
        $6p_-{\hspace{-0.5ex}}^2$ & 1.31\%\\
        $6s7s$ & 0.74\%\\
        $5d_+{\hspace{-0.5ex}}^2$ & 0.56\%\\
        $5d_-{\hspace{-0.5ex}}^2$ & 0.35\%\\ \colrule
        \multicolumn{2}{c}{${}^3P_0^o$}\\ \colrule
        $6s6p_-$ & 95.60\%\\
        $5d_-6p_+$ & 1.02\%\\
        $6p_-7s$ & 0.48\%\\ \colrule
        \multicolumn{2}{c}{${}^3P_1^o$}\\ \colrule
        $6s6p_-$ & 73.41\%\\
        $6s6p_+$ & 22.02\%\\
        $5d_-6p_+$ & 0.68\%\\
        $5d_-6p_-$ & 0.46\%\\
        $6p_-7s$ & 0.38\%
    \end{tabular}
    \end{ruledtabular}
    \label{tab:CSFs jj}
\end{table}

\section{\label{app:c} UNCERTAINTY ESTIMATES FOR ISOTOPE SHIFT PARAMETERS}
Systematic expansions of the active space and correlation model have been undertaken in order to estimate the uncertainties for the isotope shift parameters, $K$ and $F$. The error introduced by truncating the active space at $12spdf$ is estimated by analysing the $K$ and $F$ values after adding each new correlation layer. This analysis was performed using ${}^{174}\mathrm{Yb}$ with core-valence correlations restricted to single excitations from $5s,5p,4f$ and unrestricted valence-valence correlations. The results are presented in Table \ref{tab:AS_convergence}. Based on these results, the uncertainty in the final $K$ and $F$ values due to the truncated active space is estimated to be the absolute difference between the $12sp11d10f$ and $11sp10d9f$ layers, as these were the largest two correlation layers added with an orbital of each symmetry.

\begin{table}[h]
    \centering
    \caption{Sequences of isotope shift parameters upon addition of correlation layers. The final row constitutes estimates for the uncertainty in each of the parameters for each isotope. The units for $K$ are GHz\,u and the units for $F$ are GHz\,$\mathrm{fm}^{-2}$.}
    \begin{ruledtabular}
    \begin{tabular}{ccccc}
        Layer & $K_{\mathrm{clock}}$ & $F_{\mathrm{clock}}$ & $K_{\mathrm{ICL}}$ & $F_{\mathrm{ICL}}$ \\ \colrule
        $7sp6d5f$ & -192.67 & -10.4060 & -176.48 & -10.5151\\
        $8sp7d6f$ & -204.07 & -10.0544 & -189.54 & -10.1697\\
        $9sp8d7f$ & -282.05 & -10.9553 & -273.16 & -11.0695\\
        $10sp9d8f$ & -268.73 & -10.9639 & -257.58 & -11.0710\\
        $11sp10d9f$ & -290.47 & -10.8480 & -281.57 & -10.9548\\
        $12sp11d10f$ & -288.15 & -10.8386 & -279.59 & -10.9418\\
        $12spdf$ & -288.07 & -10.8393 & -279.55 & -10.9425\\ \colrule
        Uncertainty estimate & 2.4 & 0.0094 & 2.0 & 0.013
    \end{tabular}
    \end{ruledtabular}
    \label{tab:AS_convergence}
\end{table}

The error introduced by restricting the core-valence correlation to single excitations from $5s,5p,4f$ is estimated similarly, by analysing the $K$ and $F$ values with increasingly more core orbitals available for excitation. This analysis was performed using ${}^{176}\mathrm{Yb}$ with the active space up to $12spdf$ and unrestricted valence-valence correlation. The results are presented in Table \ref{tab:CV_convergence}. Based on these results, this uncertainty is estimated to be twice the absolute difference between this core and the next largest available core of $5s,4d,5p,4f$.

\begin{table}[h]
    \centering
    \caption{Sequences of isotope shift parameters upon inclusion of deeper core-valence correlation. The final row constitutes estimates for the uncertainty in each of the parameters for each isotope. The units for $K$ are GHz\,u and the units for $F$ are GHz\,$\mathrm{fm}^{-2}$.}
    \begin{ruledtabular}
    \begin{tabular}{ccccc}
        Available core & $K_{\mathrm{clock}}$ & $F_{\mathrm{clock}}$ & $K_{\mathrm{ICL}}$ & $F_{\mathrm{ICL}}$ \\ \colrule
        $4f$ & 54.90 & -9.4283 & 56.36 & -9.5023\\
        $5p,4f$ & -265.41 & -10.5830 & -255.65 & -10.6681\\
        $5s,5p,4f$ & -288.05 & -10.8326 & -279.53 & -10.9358\\
        $5s,4d,5p,4f$ & -325.22 & -10.8255 & -315.26 & -10.9297\\ \colrule
        Uncertainty estimate & 75 & 0.015 & 72 & 0.013
    \end{tabular}
    \end{ruledtabular}
    \label{tab:CV_convergence}
\end{table}

\section{\label{app:d} ALTERNATIVE PRESENTATION OF DIFFERENCES IN MEAN-SQUARE CHARGE RADII}

The differences in mean-square charge radii are presented in Table \ref{tab:delta r squared} for pairs of isotopes. Alternatively, a single reference isotope may be chosen and differences in mean-square charge radii given relative to this reference isotope. For ytterbium, this reference isotope is commonly chosen to be ${}^{176}\mathrm{Yb}$. Differences in mean-square nuclear charge radii of this type are presented in Table \ref{tab:delta r squared relative to 176Yb}, with the reference isotope of ${}^{176}\mathrm{Yb}$.

\begin{table}[h]
    \centering
    \caption{Differences in mean-square nuclear charge radii relative to ${}^{176}\mathrm{Yb}$, $\delta\langle r^2\rangle^{176,A'}$, in units of $10^{-3}\,\mathrm{fm}^2$ (column 2). Data from other works are presented for comparison.}
    \begin{ruledtabular}
    \begin{tabular}{cccc}
        $A'$ & This work & Ref. \cite{Martensson-Pendrill1994} & Ref. \cite{Angeli2013}\footnote{Ref. \cite{Angeli2013} presents only statistical errors in the uncertainty --- the large uncertainty in the mass shift parameter used in calculation is not propagated through. Propagating the uncertainty from the mass shift parameter leads to an uncertainty of $\sim9\times10^{-3}\,\mathrm{fm}^2$ for the first row.}\\ \colrule
        174 & 94.4(0.5) & 90(2) & 115.9(0.1)\\
        173 & 149.3(0.8) & 142(3) & 181.0(0.1)\\
        172 & 193.4(1.0) & 184(5) & 236.6(0.1)\\
        171 & 274.4(1.3) & 259(6) & 327.3(0.1)\\
        170 & 320.0(1.6) & 303(7) & 384.5(0.1)\\
        168 & 454.8(2.1) & 428(13) & 540.6(0.3)
    \end{tabular}
    \end{ruledtabular}
    \label{tab:delta r squared relative to 176Yb}
\end{table}

\section{\label{app:e} CLOCK TRANSITION FREQUENCIES}

Table~\ref{tab: clock freqs} lists our estimates for the absolute \clockT\ transition frequencies in neutral ytterbium for isotopes where it is yet to be measured, together with the known frequencies.  Our estimates and their uncertainties are based on the weighted mean of the isotope shift values presented in Table~\ref{tab:clock isotope shift estimates} using Methods 2 and 3, and the existing absolute transition frequency measurements.

\begin{table}[h]
    \centering
    \caption{Estimated (this work) and previously measured clock transition frequencies in Yb \textsc{i}. }
    \begin{ruledtabular}
    \begin{tabular}{ccc}
        Isotope & Transition frequency (MHz) \\ \colrule
        168 & 518 297 652.3(3.5) \\
        170 & 518 296 294.7(1.4) \\
        171 &  518 295 836.59086361(13) \cite{Pizzocaro2020}\\
        171 &  518 295 836.59086371(11) \cite{McGrew2019}\\
        172 & 518 295 019.7(1.9) \\
        173 & 518 294 576.845268(10) \cite{Clivati2016}\\
        174 & 518 294 025.3092178(9) \cite{Poli2008} \\
        176 & 518 293 076.4(2.7)
    \end{tabular}
    \end{ruledtabular}
    \label{tab: clock freqs}
\end{table}

\bibliography{refs}

\end{document}